\documentclass[tikz,11pt,english,a4paper]{article}
\usepackage{float}
\usepackage{jcappub}
\usepackage[utf8]{inputenc}

\usepackage{bm}
\usepackage{algorithm}
\usepackage{nicefrac}
\usepackage{array}
\usepackage{multirow}
\usepackage{subfigure}

\usepackage[noend]{algpseudocode}
\makeatletter
\def\BState{\State\hskip-\ALG@thistlm}
\makeatother

\usepackage{fontawesome5}
\usepackage{hyperref}
\makeatletter
\newcommand{\github}[1]{%
   \href{#1}{\faGithub}%
}
\makeatother

\usepackage{lmodern}

\newcommand{\class}{\textsc{class}}

\newcommand{\concept}{\textsc{concept}}
\newcommand{\dtfe}{\textsc{dtfe}}
\usepackage{siunitx}
\DeclareSIUnit \parsec {pc}

\usepackage[T1]{fontenc}

\usepackage{soul}

\usepackage{lmodern} 
\rmfamily 
\DeclareFontShape{T1}{lmr}{b}{sc}{<->ssub*cmr/bx/sc}{}
\DeclareFontShape{T1}{lmr}{bx}{sc}{<->ssub*cmr/bx/sc}{}

\begin{document}
\emergencystretch 3em


\title{Cosmic curl – Features and convergence of the vorticity power spectrum in $N$-body simulations}

\author[a]{Camilla T. G. Sørensen,}
\author[a]{Steen Hannestad,}
\author[a]{and Thomas Tram}

\affiliation[a]{Department of Physics and Astronomy, Aarhus University,
 DK-8000 Aarhus C, Denmark}

\emailAdd{camth@phys.au.dk}
\emailAdd{steen@phys.au.dk}
\emailAdd{thomas.tram@phys.au.dk}

\abstract{Observations of the cosmic velocity field could become an important cosmological probe in the near future. To take advantage of future velocity-flow surveys we must however have the theoretical predictions under control. In many respects, the velocity field is easier to simulate than the density field because it is less severely affected by small-scale clustering. Therefore, as we also show in this paper, a particle-mesh (PM) based simulation approach is usually sufficient, yielding results within a few percent of a corresponding $\text{P}^3\text{M}$ simulation in which short-range forces are properly accounted for, but which also carry a much larger computational cost.

However, in other respects the velocity field is much more challenging to deal with than the density field: Interpolating the velocity field onto a grid is significantly more complicated, and the vorticity field (the curl-part of the velocity field) is severely affected by both sample variance and discretisation effects. While the former can be dealt with using fixed amplitude initial conditions, the former makes it infeasible to run fully converged simulations in a cosmological volume. However, using the $N$-body code \concept{} we show that one can robustly extrapolate the cosmic vorticity power spectrum from just 4 simulations with different number of particles. We expect our extrapolated vorticity power spectra to be correct within 5\% of the fully converged result across three orders of magnitude in $k$. Finally, we have also investigated the time dependence of the vorticity as well as the ratio of vorticity to divergence.}

\maketitle

\section{Introduction}\label{sec:introduction}

Over the past two decades, large scale surveys of cosmic structure have reached a scale and precision that have made them an important tool in precision cosmology. Most surveys have focused on providing information on the cosmic matter density field, either via number counts of galaxies or via measurements of galaxy shapes which in turn allows for measurements of the matter distribution using weak gravitational lensing information.
While the use of number counts is in principle straightforward, it comes with a number of important challenges such as uncertainties about completeness, mass-luminosity bias effects, etc. Using weak lensing circumvents these problems because it allows for a direct measurement of the gravitational potential. However, weak lensing suffers from the problem of having access only to line-of-sight information and even using tomography, 3D information is somewhat difficult to extract.
Despite these various challenges, cosmological surveys such as DESI \cite{DESI:2025zgx,DESI:2024hhd} and Euclid \cite{Euclid:2024yrr} are now reaching percent level precision on the inferred cosmological matter power spectrum.

A completely different possibility is to use measurements of cosmic velocity fields to extract information about cosmic structure. If reliable distances and radial velocities can be measured for a large numbers of objects at cosmological distances, it is in principle possible to reconstruct e.g. cosmological velocity power spectra which in turn can be used to perform cosmological parameter inference. 
Velocity measurements have a number of advantages over measurements of the density field. For example they do not rely on knowing the completeness of the sample since they are inherently volume weighted rather than density weighted (see e.g. \cite{Bernardeau:2001qr,Hannestad:2007fb}). However, they have the distinct disadvantage of requiring very accurate distance measurements, i.e.\  they require a set of very good standard candles.
This limitation means that cosmic velocity measurements (beyond the important measurement of the current background expansion rate of the Universe, $H_0$) have not yet reached a level of precision where they are competitive with galaxy surveys. However, this situation is likely to change with the advent of very large scale supernova surveys from facilities such as the Vera C. Rubin telescope (see e.g.~\cite{LSST}). These will allow for the detection of hundreds of thousands of type Ia supernovae and provide a detailed 3D map of the cosmic velocity field \cite{Hannestad:2007fb,Odderskov:2016qig,Howlett:2017asw}.

With the advent of very precise cosmological surveys demands on the corresponding theoretical calculations of observables naturally increases. Current measurements of the matter power and lensing power spectra, for example, require that theoretical computations are accurate at percent level precision over a wide range of scales. This is generally achievable using sets of $N$-body simulations which include effects from the baryonic component such as star formation and supernova feedback
(see e.g.~\cite{Schneider:2015yka} for a discussion).
However, the precision with which observables related to velocity can be calculated are much less well known. In particular, the generation of vorticity in the cosmic velocity field could be problematic because it is not sourced at lowest order in perturbation theory and is therefore only generated by the emergence of non-linear structures
(see e.g.~\cite{Pueblas:2008uv,Gordon:2007zw,Umeh:2023lbc,Jelic-Cizmek:2018gdp,Garny:2022tlk,Garny:2025zlq,Garny:2025ovs}).
In this paper we provide a detailed investigation of vorticity generation in cosmological $N$-body simulations with particular emphasis on convergence properties. Section 2 provides an overview of the required theory, and section 3 describes our numerical simulation setup. Section 4 discusses how to best extract velocity divergence and vorticity spectra from simulations by means of Delauney tesselation. We present our main results in section 5, and finally we provide a discussion and conclusion in section 6.

\section{Theory}\label{sec:theory}

Analysing cosmic velocity fields, both from numerical $N$-body simulations and from real measurements, is done by calculating the power spectra for the velocity divergence, $\nabla \cdot \vec{v}$, and the velocity vorticity, $\nabla \times \vec{v}$. These power spectra can then be used to obtain information about the structure of the velocity field at different physical scales \cite{vorticity}.

The divergence $\theta$ is defined as the gradient of the velocity field

\begin{equation}
	\theta = \nabla \cdot \vec{v} = \frac{\partial v_i}{\partial x_i} + \frac{\partial v_j}{\partial x_j} + \frac{\partial v_k}{\partial x_k}
	\label{eq:div}
\end{equation}

and the vorticity $\vec{\omega}$ is defined as the curl of the velocity field $\omega^k = \epsilon^{ijk} \frac{\partial v_j}{\partial x_i} $  where $\epsilon^{ijk}$ is the Levi-Civita symbol. We find it convenient to define the quantity

\begin{equation}
	\omega_{ij} = \frac{1}{2} \left[ \frac{\partial v_i}{\partial x_j} - \frac{\partial v_j}{\partial x_i} \right] \,,
	\label{eq:vor}
\end{equation}

primarily because it coincides with the output of the numerical code \dtfe{} which we shall introduce later.

After the divergence and vorticity for the velocity field has been calculated, the power spectra can be calculated in the usual way. Divergence is a scalar quantity, so $P_\theta$ is constructed directly from $\theta$. The vorticity vector has three components that due to isotropy should be statistically identical. Thus, we compute the power spectra of each component independently and add the power spectra together. In summary, 

\begin{equation}
	P_\theta = \text{Power spectra of } \theta \,,
	\label{eq:div_p}
\end{equation}

\begin{equation}
	P_\omega = P_{\omega_{12}} + P_{\omega_{13}} + P_{\omega_{23}} \,,
	\label{eq:vor_p}
\end{equation}

where the factor of $\frac{1}{2}$ in equation~\eqref{eq:vor} ensures that this matches the usual definition of $P_\omega$.
The cosmological model used throughout this work is a spacially flat $\Lambda$CDM model compatible with the Planck 2015 result as well as reference~\cite{Jelic-Cizmek:2018gdp}: $h = 0.67556$, $\Omega_\text{b}h^2 = 0.022032$, $\Omega_\text{CDM}h^2 = 0.12038$, $A_\text{s} = 2.215 \cdot 10^{-9}$, $n_\text{s} = 0.9619$, and $k_* = 0.05 \text{ Mpc}^{-1}$.

It should be noted that the initial conditions are set using linear perturbation theory where $\vec{\omega} = 0$, but because of discretisation and finite numerical precision, some vorticity will be generated initially.

\section{Numerical simulation setup}\label{sec:simulations}
We have used the $N$-body code \concept{}~\cite{Dakin:2021ivb} to make all $N$-body simulations used in this work. \concept{} is an $N$-body code designed for simulations of large-scale structure formation. It utilises the Einstein–Boltzmann solver code \class{}~\cite{Blas:2011rf} to generate the initial conditions using linear perturbation theory. The simulation of large-scale structure formation is done using a P$^3$M gravity solver, which uses the PM method on larges scales augmented by short-ranged direct summation to make it possible to get precise results at both small and large scales. Simulations in \concept{} can also be run using only the PM method, which is orders of magnitudes faster, but does not resolve the small scale structure because of the lack of short-range forces. All simulations done with \concept{} have been run with the physical model given in section~\ref{sec:theory}, and the simulation outputs are snapshots of the particles positions and velocities at different scale factors in the snapshot format of \textsc{gadget 2} \cite{Springel:2005mi}.

Since the outputs from \concept{} are snapshots of the \textsc{gadget 2} format, the unit for the length is kpc/h, and the unit for the velocity is in km/s. Furthermore, the velocity saved in the \textsc{gadget 2} snapshot is $\frac{1}{\sqrt{a}} \cdot \dot{\chi}$, where $a$ is the scale factor, and $\dot{\chi}$ is the comoving velocity with $v = a \cdot \dot{\chi}$. The physical velocity $v$ is therefore found by multiplying the velocity from the snapshot with the factor $\sqrt{a}$.

\subsection{PM vs P$^3$M}
\label{sec:pmvsp3m}
For studies of cosmological density fields it is well known that PM-based simulations significantly underestimate power on small (very non-linear) scales because of their inability to properly resolve short range forces. However, for the purpose of studying velocity fields the a priori expectation is that this effect is substantially less dramatic because velocity fields are dominated by large scale density fluctuations rather than strongly varying local density fluctuations. This can be argued based on the fact that the local potential can be obtained via spatial integration of the Poisson equation, $\nabla^2 \Phi = 4 \pi G \rho_0 \delta$ and the acceleration subsequently obtained from the gradient of the potential via $\dot{\vec{v}} = - \nabla \Phi$, leading to $\vec{v}$ being substantially smoother than $\delta$.
Therefore it seems likely that PM simulations should be able to capture all the relevant features of cosmic velocity fields, potentially leading to a much reduced computation time.

\begin{figure}
	\centering
	\subfigure[]{\includegraphics[width = 0.49\textwidth]{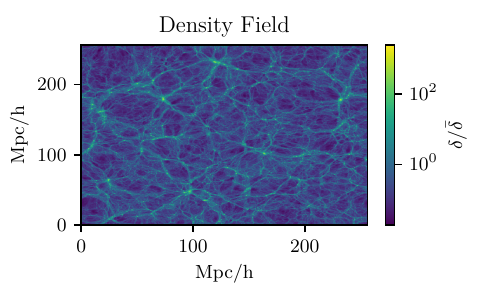}}
	\subfigure[]{\includegraphics[width = 0.49\textwidth]{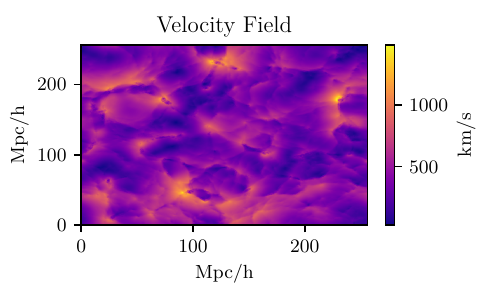}}
	\caption{(a) The density field plotted in log-scale for $z = 0$ for the simulation with $N = 1024^3$ particles, $n_\text{grid} = 2048$, and $L_\text{box} = 256$ Mpc/h using the P$^3$M method. (b) The velocity field for the same simulation plotted in linear scale. Note that the variation in the velocity field are much more linear than the variation in the density field.}
	\label{fig:2Dslice}
\end{figure}

We have investigated this possibility by comparing P$^3$M and PM simulations for a setup with $N = 1024^3$ particles, a grid size of $n_\text{grid} = 2048$ \footnote{$n_\text{grid} = 2\sqrt[3]{N}$ is the default choice in \concept{} because it strikes a good balance between increased memory requirements with larger $n_\text{grid}$ and increased CPU requirements with smaller $n_\text{grid}$.}, and a box size of $L_\text{box}$ = 256 Mpc/h. P$^3$M and PM simulations are run with identical random amplitudes and phases to remove realisation effects when comparing power spectra. From figure~\ref{fig:2Dslice} it can be clearly seen that the velocity field is substantially more linear in nature than the corresponding density field. 

\begin{figure}
	\centering
	\subfigure[]{\includegraphics[width = 0.49\textwidth]{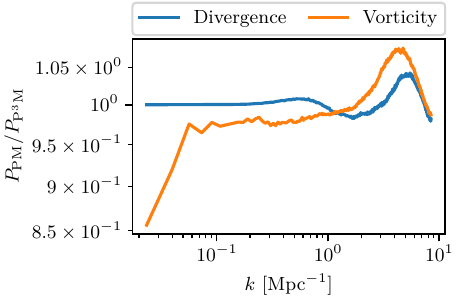}}
	\subfigure[]{\includegraphics[width = 0.49\textwidth]{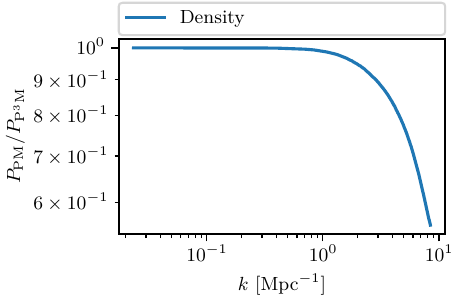}}
	\caption{(a) $\frac{P_\text{PM}}{P_{\text{P}^3\text{M}}}$ plotted for the power spectra of the velocity divergence and vorticity respectively. (b) $\frac{P_\text{PM}}{P_{\text{P}^3\text{M}}}$ plotted for the power spectra for the density field. It can be seen that the power spectra for the velocity divergence and vorticity for PM and P$^3$M respectively agree with each other within ten percent even on small scales. However, the power spectra for the density field for PM and P$^3$M respectively do not agree at all on small scales.}
	\label{fig:p3mvspm}
\end{figure}

Next, in figure~\ref{fig:p3mvspm} we plot the ratios of various power spectra between P$^3$M and PM simulations. It can be clearly seen that for the density power spectra the PM simulation indeed severely underestimates small scale power. However, both divergence and vorticity power spectra are identical to within approximately 5–10\% precision, justifying the use of PM simulations when studying cosmic flow velocities.
Note that the method used for extracting the velocity divergence and vorticity power spectra is the velocity gradient method discussed in the next section.

Having established that the PM method is sufficiently accurate on all relevant scales for computing velocity power spectra, we will use \concept{} in PM mode for all subsequent simulations related to this work.

\section{Methods for extracting velocity power spectra}\label{sec:velspec}

There are well-established standard methods available for calculating density power spectra from $N$-body simulations. The typical pipeline is to first use cloud-in-cell (CIC) interpolation of the particle list onto a Cartesian grid in real space and then use FFT techniques to produce the power spectrum of density fluctuations, including a deconvolution of the CIC kernel.
This method works well, independently of whether local particle densities are low or high, and the calculation of power spectra is unproblematic even in the case of empty grid cells. However, for velocity power spectra this is no longer the case. The estimated velocity field should be smooth at all points in space, but if an estimate of the velocity field is attempted by interpolating particle velocity components onto a regular Cartesian grid, the appearance of empty cells will lead to a breakdown of this property: Empty cells in the resulting grid have ill defined velocities.
This fundamental problem can in principle be remedied by increasing the number of particles in the simulation relative to the grid size, but since structures tend to become very non-linear at late times this method becomes very inefficient~\cite{Hannestad:2007fb}.


To get around this inefficiency of the CIC interpolation, several other methods have been proposed. One option is to use smoothed particle hydrodynamics (SPH) estimators (e.g.~\cite{Hernquist:1988zk}), but they tend to introduce excessive smoothing which, in contrast to CIC interpolation, cannot be analytically removed from the power spectrum. Another idea~\cite{Schaap:2000se} is to use Delaunay interpolation which is a standard method for interpolating data known on an unstructured grid. This method is commonly referred to as Delaynay Tesselation Field Estimator (DTFE). However, DTFE may lead to excess power in coarse-grained derivative-properties like divergence and vorticity in multi-stream regions as pointed out by Refs.~\cite{Abel:2011ui, Hahn:2014lca}. These authors then proposed to take the \emph{initial} configuration of particles as vertices in the Delaunay tetrahedra which could then be used at any later stage to interpolate the phase-space distribution function. This phase-space method is better at resolving multi-stream regions at the cost of a somewhat worse performance in single-stream regions compared to the DTFE method~\cite{Feldbrugge:2024wcm}. A combination of the latter two methods, denoted PS-DTFE, was very recently proposed~\cite{Feldbrugge:2024wcm}.

Another strategy is to use CIC together with some method for dealing with empty cells. Ref.~\cite{Jelic-Cizmek:2018gdp} proposed various ways to fill an empty cell by using the last known value of the velocity in the given cell, while Ref.~\cite{Garny:2022kbk} introduce a novel multigrid method where empty cells are filled based on coarser grids. In this work we use the DTFE method as implemented in the code \textit{The Delaunay Tesselation Field Estimator code} (\dtfe{})~\cite{vorticity, Cautun:2011gf}. \dtfe{} takes a snapshot as input, and it can output grids for e.g. the volume averaged density, the volumed averaged velocity, the volumed averaged velocity gradients etc. Using \dtfe{} directly on large simulations is not straightforward though as described in appendix~\ref{sec:dtfemod}. Our modified fork of \dtfe{} is available at \url{https://github.com/AarhusCosmology/DTFE}.

\subsection{Calculating the vorticity power spectrum}\label{sec:vorcompare}
\begin{figure}
	\centering
	\subfigure[]{\includegraphics[width = 0.49\textwidth]{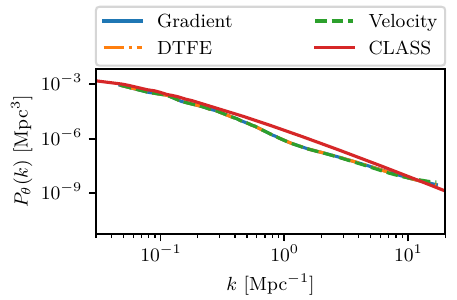}}
	\subfigure[]{\includegraphics[width = 0.49\textwidth]{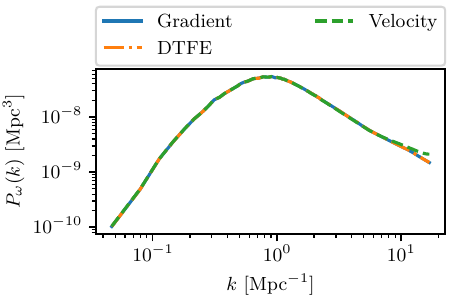}}
	\caption{(a) The divergence for the simulation with $N = 1024^3$ particles, $n_\text{grid} = 1024$, and $L_\text{box} = 128\, \text{Mpc}/h$ using the PM method. 'Gradient' and 'DTFE' is the divergence calculated with the velocity gradient method using the gradient and divergence as \dtfe{} outputs respectively. 'Velocity' is the divergence calculated with the velocity field method, and 'CLASS' is the theoretical divergence power spectrum calculated with \class{}. (b) Same simulation and methods, but with the vorticity plotted instead. Note that since the vorticity is non-linear, it cannot be calculated theoretically using \class{}.}
	\label{fig:construction}
\end{figure}

There are two methods to calculated the divergence and vorticity power spectra using three different outputs from \dtfe{}. All methods use the function \textsc{FFTPower} form the \textsc{Python} package \textsc{nbodykit}~\cite{Hand:2017pqn} to calculate the power spectra. \textsc{FFTPower} takes a mesh grid from real space as input, and gives the power spectrum and associated $k$-values as output. The power spectra found from both methods can be seen in figure~\ref{fig:construction}. \\

\noindent \textbf{Velocity method}
The divergence and vorticity power spectra can be found from a grid of the volume averaged velocity by first transforming the velocity grid into fourier space, and then calculating the velocity gradient and vorticity, remembering that taking a derivative of the velocity in fourier space is the same as multiplying the velocity with $ik$. The divergence becomes

\begin{equation}
	\theta(\vec{k}) = i (k_1 v_1 + k_2 v_2 + k_3 v_3) \,,
\end{equation}
and the vorticity becomes
\begin{equation}
	\omega_{ij}(\vec{k}) = \frac{1}{2} i (k_i v_j - k_j v_i) \,.
\end{equation}

When $\theta_k$  and all three independent $\omega_{ij}$ are calculated, they are transformed back into real space, and the power spectra are then calculated using equations~(\ref{eq:div_p}–\ref{eq:vor_p}) and \textsc{FFTPower}. \\

\noindent \textbf{Velocity gradient method}
The velocity divergence and vorticity can be found from a grid of the volumed averaged velocity gradient in two ways. The first is by getting the volume averaged field from \dtfe{}, and then calculating the divergence and velocity by using equations~\eqref{eq:div} and~\eqref{eq:vor} respectively. The second way is to get the divergence and velocity power spectra directly from \dtfe{}. To show that these two methods can be used interchangeably, we plot both in figure~\ref{fig:construction}. 


To test all three methods, the velocity divergence and vorticity have been computed using all three methods for a simulation with $N = 1024^3$ particles, $n_\text{grid} = 1024$, and $L_\text{box} = 128$ Mpc/h using the PM method. The PM method is used because it is shown in section~\ref{sec:pmvsp3m} that it gives results comparable to the P$^3$M method while being a lot less computational expensive to do\footnote{A comparison of the three methods have also been done using the P$^3$M method, and it leads to the same conclusion.}. The result can be seen in figure~\ref{fig:construction}. For the velocity divergence, the theoretical linear power spectrum has also been calculated using \class{} and plotted. As it can be seen from the figure, the two methods give the same power spectra, but the velocity gradient method is slightly better for large $k$-values. Furthermore, the gradient method is faster to use than the velocity method, and only the method using the velocity gradient will therefore be used from now on.

\section{Numerical results}


Having established our preferred method for simulating and extracting cosmological vorticity power spectra we proceed to perform a number of simulations to test properties of the vorticity power spectrum. These properties includes dependence on the initial seed and initial redshift as well as the numerical convergence as a function of the number of particles and box size.

\subsection{Dependence on initial seed}

It is well known that cosmological $N$-body simulations depend on the initial random seed used to generate the initial condition for the simulation. This is particularly true in cases where the simulation volume is small because non-linear evolution amplifies initially small differences in the initial condition.
In the case of the density power spectrum this effect can to a very good approximation be removed by using paired and fixed initial conditions in the simulation~\cite{Angulo:2016hjd}. The ensemble mean of simulations converges on the result obtained when running two simulations, both with fluctuation amplitudes fixed to their expectation value, and with opposite phases (i.e.\ with the phases of the second simulation chosen such that $\theta_{2,i} = \theta_{1,i}+\pi$). In fact, for the power spectrum it is typically enough to run a single simulation with fixed amplitude, whereas for higher order statistics the phase is also important~\cite{Angulo:2016hjd}.

\begin{figure}
	\centering
	\includegraphics[width = 0.49\textwidth]{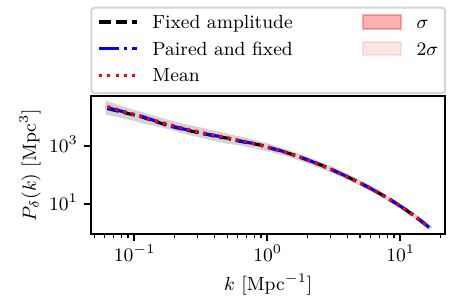}
	\includegraphics[width = 0.49\textwidth]{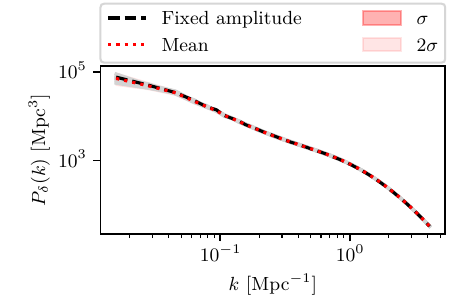}
	\includegraphics[width = 0.49\textwidth]{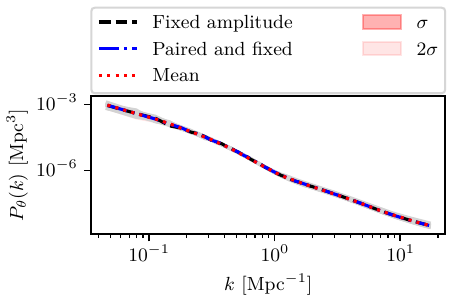}
	\includegraphics[width = 0.49\textwidth]{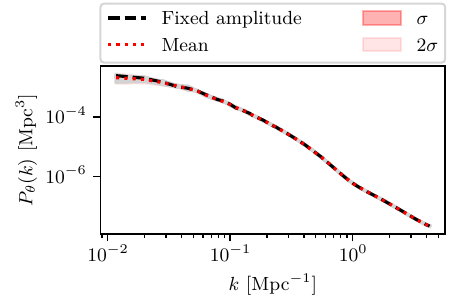}
	\includegraphics[width = 0.49\textwidth]{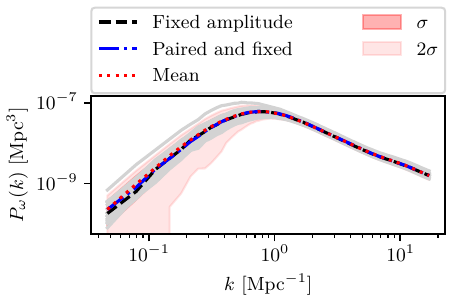}
	\includegraphics[width = 0.49\textwidth]{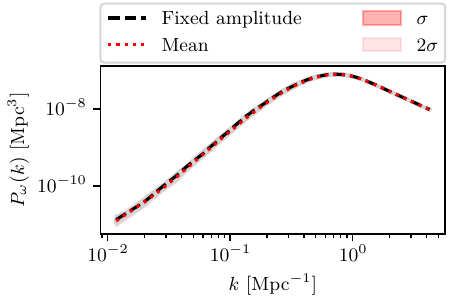}
	\caption{\label{fig:variance}Variance of the density, divergence and vorticity power spectra at redshift $z=0$. \emph{Left panels}: $128\,\text{Mpc/h}$ box. \emph{Right panels}: $512\,\text{Mpc/h}$ box. The vorticity power spectra are severely affected by sampling variance across a wide range of scales, especially in the $128\,\text{Mpc/h}$ box. Even in this case, however, the fixed amplitude simulation is consistent with the mean value.}
\end{figure}

In order to test this property on velocity power spectra we have run a set of 10 simulations in a $512 \, \text{Mpc/h}$ box and a set of 20 simulations in a $128\, \text{Mpc/h}$ box, and the results are shown in figure~\ref{fig:variance}.

Not surprisingly, we can see that all spectra have significantly larger variance in the smaller box because of the amplification effect described above. Importantly, we also see that in the 128 Mpc/h simulations the vorticity has dramatically larger variance than density or divergence because it is sourced only by non-linear gravity and does not exist at the linear level. In the larger box, vorticity has a variance more similar to density and divergence, consistent with the largest scales in these boxes still being linear.

\begin{figure}
	\centering
	\includegraphics[width = \textwidth]{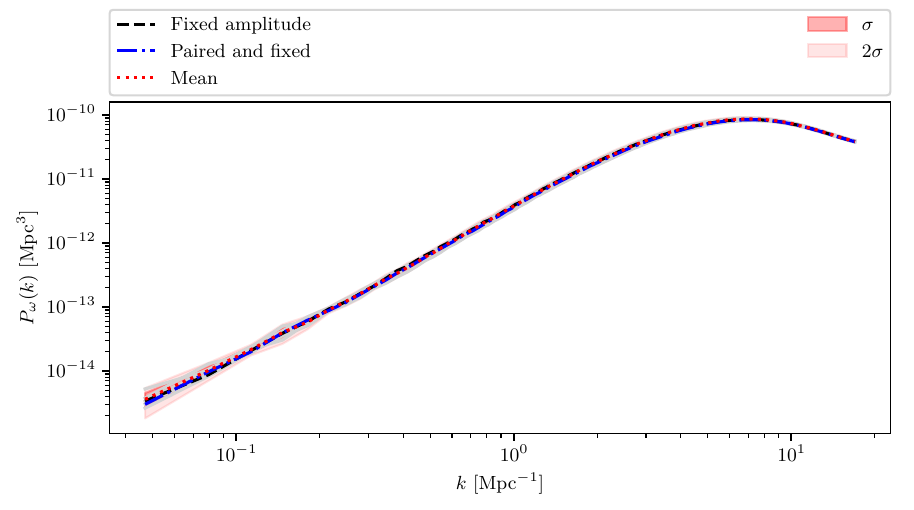}
	\caption{The variance for 5 simulations at $z = 5$ in the box with size 128 Mpc/h. It can be seen that the variance is much smaller here leading to the conclusion that the large variance in the 128 Mpc/h box at $z = 0$ is caused by strong non-linearity. As with the simulations at $z = 0$, the fixed amplitude simulation is consistent with the mean value.}
	\label{fig:variancez5}
\end{figure}

We now turn to the hypothesis that the large variance of vorticity in the small box simulations is indeed caused by strong non-linearity forming in the density field. In figure~\ref{fig:variancez5} we show the result of 5 different simulations in a $128\,\text{Mpc/h}$ box at redshift $z=5$. As conjectured, the variance is indeed much smaller than at $z=0$, and indeed 
the comoving wavenumber corresponding to the size of the box, $k_b \sim \frac{2\pi}{128 \, {\rm Mpc}} \sim 0.05 \, {\rm Mpc}^{-1}$ corresponds well to the scales going non-linear at $z=0$, but which are still linear at $z=5$.
Interestingly, this finding also means that the measured vorticity in a single volume of this size is likely to be dominated by variance.

\subsection{Dependence on initialisation redshift}

\begin{figure}
	\centering
	\includegraphics[width = 0.49\textwidth]{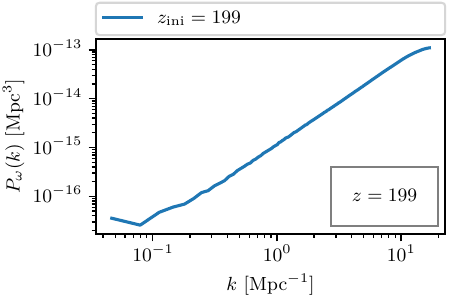}
	\includegraphics[width = 0.49\textwidth]{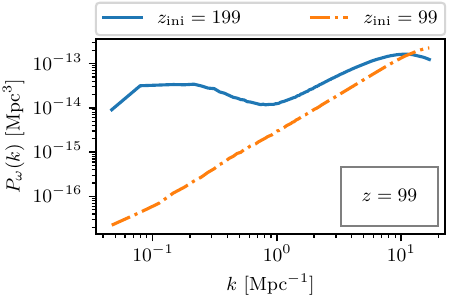}
	\includegraphics[width = 0.49\textwidth]{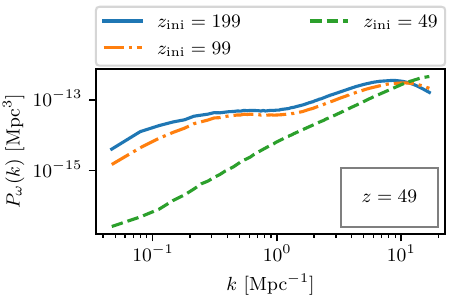}
	\includegraphics[width = 0.49\textwidth]{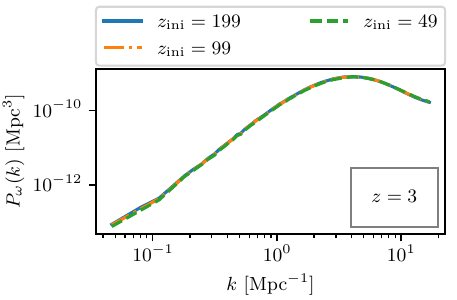}
	\includegraphics[width = 0.49\textwidth]{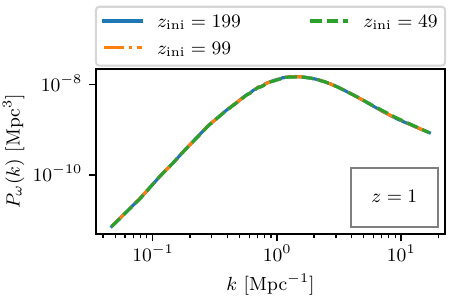}
	\includegraphics[width = 0.49\textwidth]{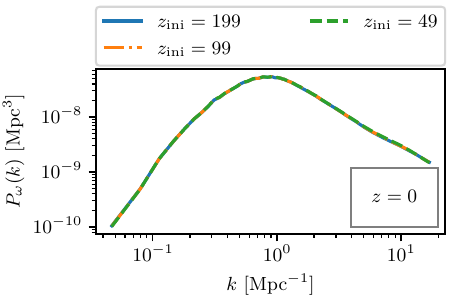}
	\caption{\label{fig:zbegin} Vorticity power spectra for different redshifts. Since the vorticity is not generated in the initial condition there is a transient build-up of power. However, the power spectra at $z=3$ is virtually indistinguishable independent of starting redshift.}
\end{figure}

Next, we have investigated how the initialisation redshift influences the late-time behaviour of the vorticity field. To this end we have run 3 different simulations with $N = 1024^3$, $n_\text{grid} = 1024$, and $L_\text{box} = 128 \text{ Mpc/h}$, starting at redshifts $z_i$= 199, 99, and 49 respectively. The results of these three runs are shown in figure~\ref{fig:zbegin} at redshifts $z$= 199, 99, 49, 3, 1, and 0. From the figures it can be seen that vorticity looks dramatically different at $z=49$, depending on the initialisation redshift (but also that it has exceedingly small amplitude). However, at $z=3$ this is reduced to a minute difference on the largest scales for the $z_i=49$ run, and at smaller redshifts the difference is negligible. This demonstrates that a starting redshift of $z_i=99$ produces robust results at the redshifts where vorticity might plausibly be observable.

\subsection{Numerical convergence}
\label{sec:convergence}
\begin{figure}
	\centering
	\includegraphics[width = 0.95\textwidth]{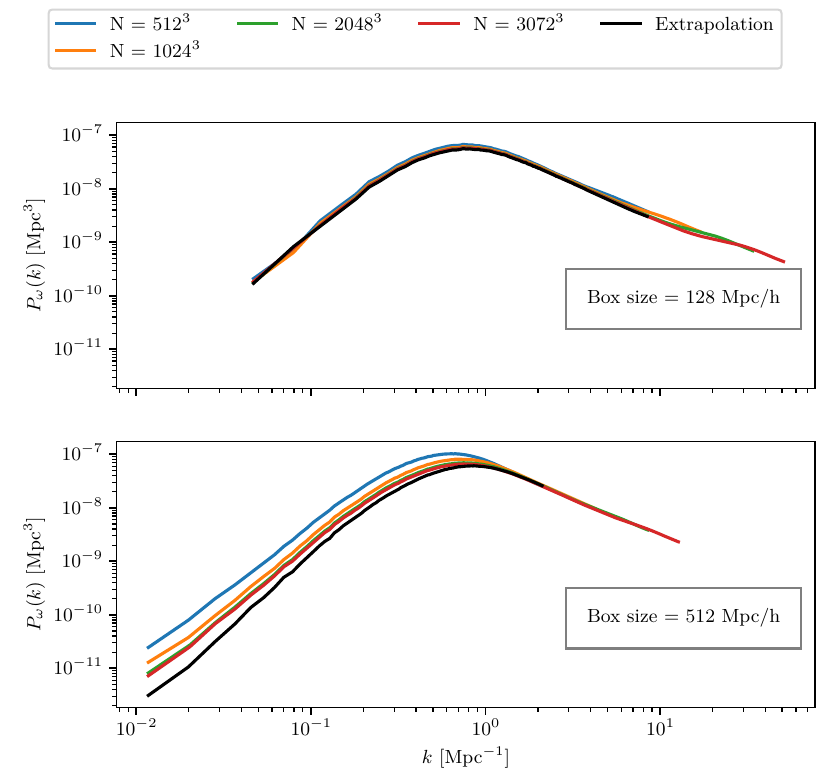}
	\caption{\label{fig:Pk_varying_N_log}Vorticity power spectra with varying number of particles for a box size of $L_\text{box} = 128\,\text{Mpc/h}$ \emph{(upper panel)} and  $L_\text{box} = 512\,\text{Mpc/h}$ \emph{(lower panel)}. Convergence with respect to number of particles can be quite slow. See text for details on how we extrapolate.}
\end{figure}

\begin{figure}
	\centering
	\includegraphics[width = 0.95\textwidth]{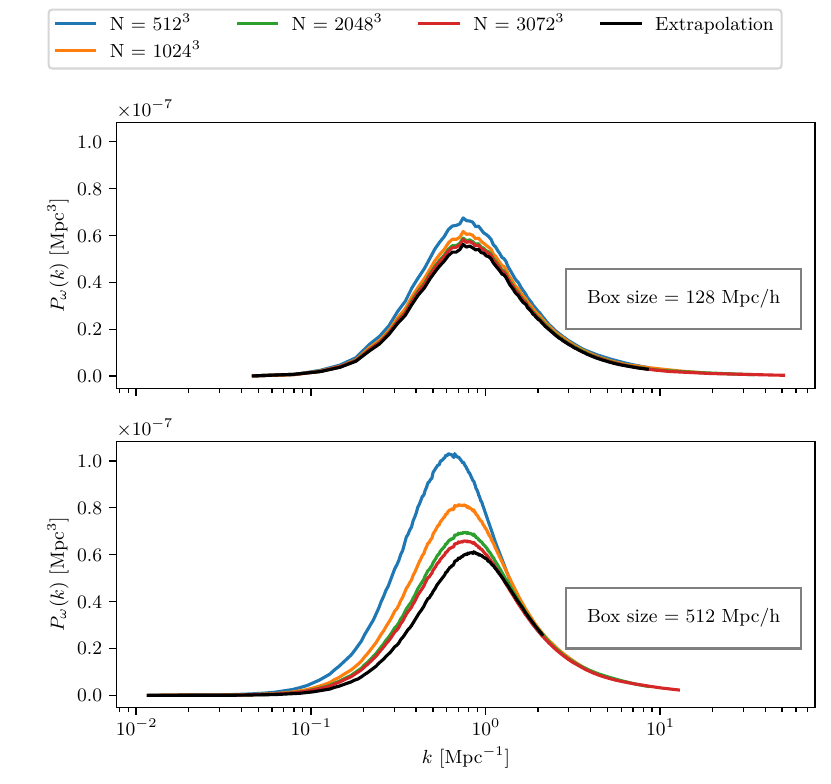}
	\caption{\label{fig:Pk_varying_N}Same as figure~\ref{fig:Pk_varying_N_log}, but with a linear y-scale. The $k$-dependent extrapolation moves the peak further to the right in the $512\,\text{Mpc/h}$ box where it coincides with the extrapolated peak in the $128\,\text{Mpc/h}$ box .}
\end{figure}

Having established a baseline for running simulations, i.e.\ initialising at a redshift of 99 and running simulations with initial amplitudes fixed to their expectation value, we proceed to test how the vorticity power spectrum depends on box size and particle number in the simulation. We ran simulations with $N=512^3, 1024^3, 2048^3\text{ and }3072^3$ particles with box sizes $L_\text{box} = \{128, 512\}\,\text{Mpc/h}$. Following Ref.~\cite{Garny:2022kbk} we introduce the parameter $X\equiv N^{\frac{1}{3}} \frac{\text{Mpc/h}}{L_\text{box}}$ which is expected to control numerical convergence.\footnote{Ref.~\cite{Garny:2022kbk} considered the mean particle separation $\lambda\equiv  L_\text{box} N^{-\frac{1}{3}}$ and the scale $k_\text{peak}$ where the vorticity power spectrum peaks. Our definition of $X$ is such that $X= (\lambda k_\text{peak})^{-1}$ with $k_\text{peak} = 1 \text{h}/\text{Mpc}$.} Looking at figure~\ref{fig:Pk_varying_N_log} we can see by eye that the $512\,\text{Mpc/h}$ box is not even converged using $3072^3$ particles while the  $128\,\text{Mpc/h}$ box seems converged. However, switching to linear space in figure~\ref{fig:Pk_varying_N} reveals that even $X=24$, $3072^3$ particles in a  $128\,\text{Mpc/h}$ box, is not enough to reach full convergence.

\begin{figure}
	\centering
	\includegraphics[width = \textwidth]{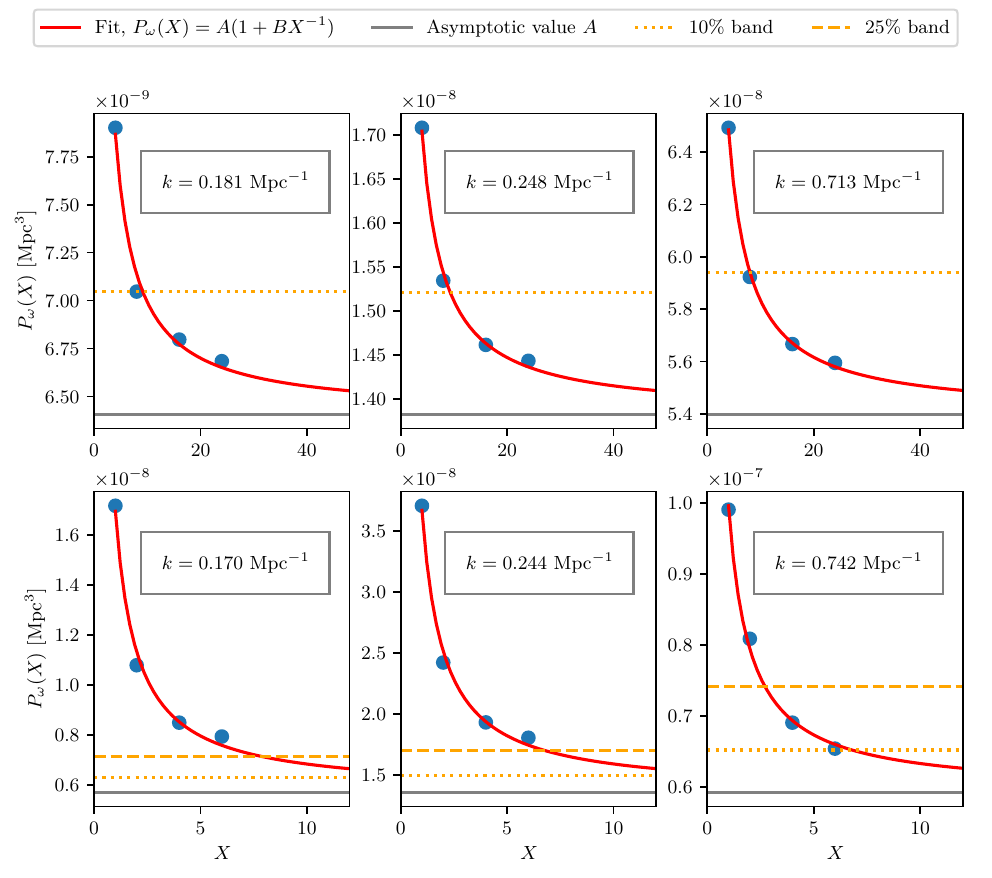}
	\caption{\label{fig:Pk_fits}A selection of the power law fits to the vorticity power spectrum. \emph{Upper panels}: 128 Mpc/h box. \emph{Lower panels}: 512 Mpc/h. The rather slow $X^{-1}$ convergence means that the asymptotic value is somewhat lower than one might expect without the fit. Note that  the $N=3072^3$ simulation is not even within the 25\% band of the asymptotic value in the $512\,\text{Mpc/h}$ box for $k\lesssim 0.5\, \text{Mpc}^{-1}$. }
\end{figure}

Given that we cannot run fully converged simulations, we shall try to recover the asymptotic value by extrapolating our power spectra for each box size to $X\rightarrow \infty$. Then, if both the $128\,\text{Mpc/h}$ box and the $512\,\text{Mpc/h}$ box yield a consistent extrapolated value, we will consider the extrapolation to be consistent. We first looked at the behaviour of $P_\omega(k, X)$ for several different $k$-values and established that convergence is proportional to $X^{-1}$. We then made a fit to each k-mode independently using a power law

\begin{align}\label{eq:powerlaw_fit}
P_\omega = A\left(1 + B X^{-1} \right) \,,
\end{align}

and inspected each of the 254 resulting fits. We restricted the $k$-range to the Nyquist frequency of the $N=512^3$ simulations in order to get 4 points in the fit, and in figure~\ref{fig:Pk_fits} we show the fits for three different $k$-values in each box. 

For large $k$-values we are however affected by an interpolation artefact. This error affects modes larger than $k \gtrsim k_\text{Nyquist}/6$, and we checked that a similar error occurs in the density power spectrum when interpolated by \dtfe{}. The $N=3072^3$ simulations are always unaffected since $k_\text{Nyquist}^{512} = k_\text{Nyquist}^{3072}/6$. However, the \emph{fitting} will be affected somewhat since the power law cannot capture this extra power present in the first few points. The effect on the asymptotic value is not significant, however.

\begin{figure}
	\centering
	\includegraphics[width = 0.95\textwidth]{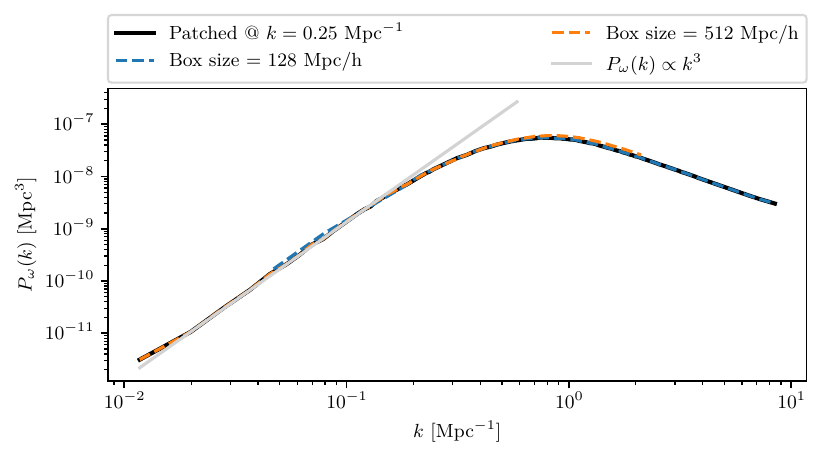}
	\caption{\label{fig:Pk_asymptotic_value}The extrapolated values for the power spectra. The two extrapolated power spectra agree for a wide range of values, and they have been patched together at $k=0.25 \, \text{Mpc}^{-1}$. On large scales the vorticity power spectrum grows approximately as $k^3$.}
\end{figure}

From equation \eqref{eq:powerlaw_fit} we have that $P_\omega \rightarrow A$ for $X \rightarrow \infty$, so the asymptotic value at each $k$ is given by $A$. In figure~\ref{fig:Pk_asymptotic_value} we have shown the extrapolated values for each box size, and we see that they overlap on a wide range of scales. This is a non-trivial result, and it shows that our extrapolation scheme yields a reasonable approximation to the fully converged result. We are patching the extrapolated power spectra at $k=0.25 \, \text{Mpc}^{-1}$ and use the extrapolated values from the $512\,\text{Mpc/h}$ box for $k < 0.25 \, \text{Mpc}^{-1}$ and the extrapolated values from the $128\,\text{Mpc/h}$ box for smaller scales.

\begin{figure}
	\centering
	\includegraphics[width = 0.95\textwidth]{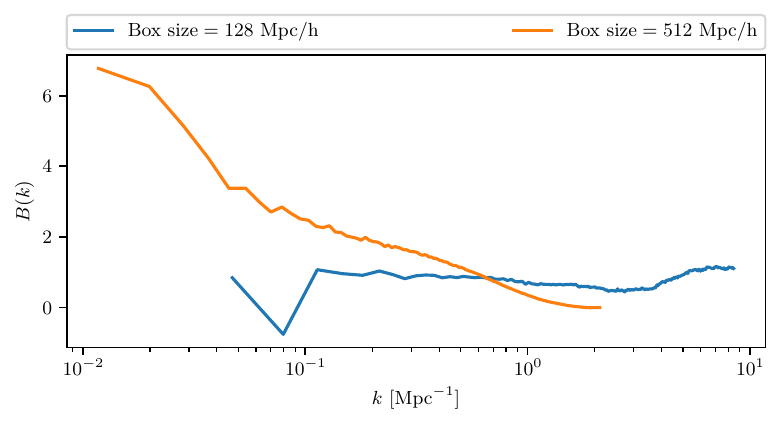}
	\caption{\label{fig:B_coefficient}The $B$ coefficients for the power law fit in equation~\eqref{eq:powerlaw_fit}. For the $128\,\text{Mpc/h}$ we find $B(k) \approx 1$ while the $512\,\text{Mpc/h}$ box has some $k$-dependence. This could be due to some mode-coupling in the larger box amplifying the error on larger scales.}
\end{figure}

There is a small discrepancy at small scales where the asymptotic value for the $512\,\text{Mpc/h}$ box is somewhat higher than the one for the $128\,\text{Mpc/h}$ box. For some reason, the $N=512^3$ power spectra is consistently \emph{lower} than the $N=1024^3$ power spectra, and this means that the fit returns an asymptotic value which is higher than the $N=3072^3$ value. We can see this explicitly in figure~\ref{fig:B_coefficient} where we have plotted the value of the $B$ coefficients. In the $128\,\text{Mpc/h}$ box we have $B \approx 1$ for all modes, but for the $512\,\text{Mpc/h}$ box it drops from a high value and becomes slightly negative for large $k$-values. Given that the asymptotic values already overlap on a wide range of scales we are not too worried about this.

Ref.~\cite{Garny:2022kbk} estimated that convergence should be achieved for $X \gtrsim 4$ based on previous $N$-body simulations. As we can see in figure~\ref{fig:Pk_fits}, $X=4$ corresponds to approximately 20\% error while $X=24$ brings the error down to the 5\% level. Based on these figures, we think the error of our extrapolated vorticity power spectrum is at most 5\%.

\subsection{Time dependence of the vorticity power spectrum}

Having established a robust method for computing the vorticity power spectrum for $z=0$ and its numerical convergence, another interesting thing to look at is the time dependence of the vorticity power spectrum.

To do this we follow the conjecture put forward in  Ref.~\cite{Pueblas:2008uv} and assume that the vorticity power spectrum can be written as a constant, $C$, times the linear growth factor to some power $\beta$ (called $n_\omega$ in Ref.~\cite{Pueblas:2008uv}), i.e.\

\begin{equation}
	P_\omega(k,z) = C \cdot (D_+(z))^{\beta(k)},
	\label{eq:P(k,z)}
\end{equation}
where $\beta(k)$ takes into account that $\beta$ might be $k$ dependent. Using the ratio $P_\omega(k,z=0)/P_\omega(k,z=1)$, $\beta(k)$ can be found as
\begin{equation}
	\beta(k) = \frac{\ln(P_\omega(k,z=0))-\ln(P_\omega(k,z=1))}{\ln(D_+(z=0))-\ln(D_+(z=1))} = -\frac{\ln(P_\omega(k,z=0))-\ln(P_\omega(k,z=1))}{\ln(D_+(z=1))},
	\label{eq:beta}
\end{equation}
and $\beta(k)$ therefore maps the time dependence of the vorticity power spectrum from $z=1$ to $z=0$.

In order to calculate $\beta(k)$ it is necessary to first calculate the vorticity power spectrum for $z=1$, and this has been done using the same approach as section \ref{sec:convergence}. 

Using the same extrapolation method as for $z=0$ turns out to be significantly harder at $z=1$ because the artificial vorticity for small $k$ still contribute significantly, particularly for the $512^3$ particles simulation in the large 512 Mpc/h box.
Our best estimate of the extrapolated $z=1$ power spectrum is shown in figure \ref{fig:P,z=1}, but we note that the result is far less robust than the corresponding $z=0$ result.
From the figure it can be seen that the power spectrum for small $k$-values approximately follows a slope proportional to $k^{2.5}$, i.e.\ slightly less than the slope of $k^3$ found for $z=0$.

\begin{figure}
	\centering
	\includegraphics[width=0.95\textwidth]{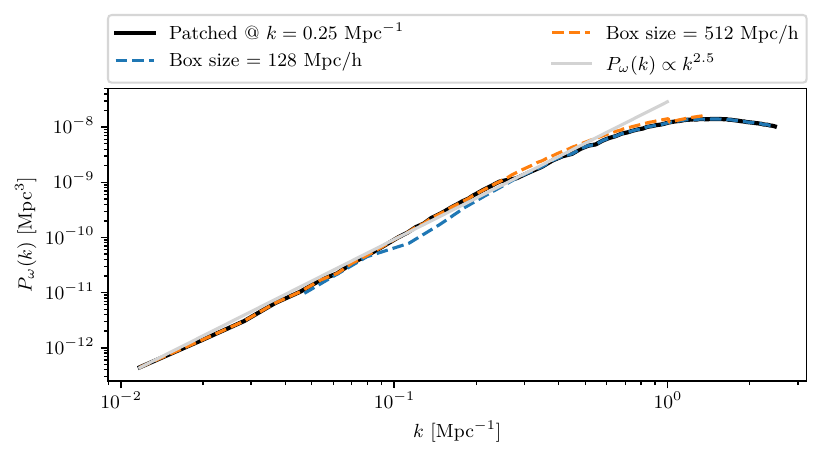}
	\caption{\label{fig:P,z=1}The extrapolated vorticity power spectrum for $z=1$. The two extrapolated power spectra have been patched together at $k = 0.25 \text{ Mpc}^{-1}$. At large scales the power spectrum grows approximately as $k^{2.5}$.}
\end{figure}

\begin{figure}
	\centering
	\includegraphics[width=1.00\textwidth]{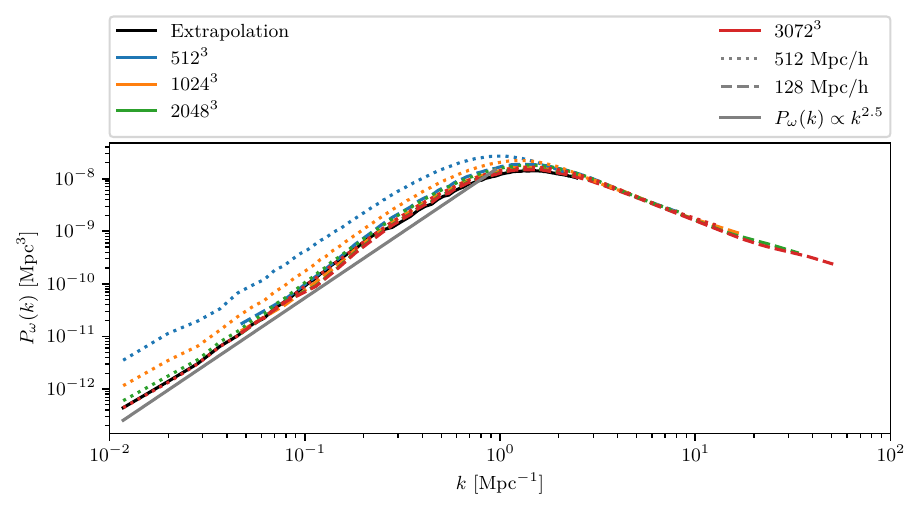}
	\caption{\label{fig:unext} The $k^{2.5}$ slope for small $k$-values plotted together with the un-extrapolated power spectra for $z=1$.}
\end{figure}

Because of the bad convergence of the $512^3$ particles simulation in the 512 Mpc/h box one might worry that the extrapolated slope of $k^{2.5}$ at small $k$-values might not be correct, and this slope is therefore plotted together with the raw $3072^3$ power spectra in figure \ref{fig:unext}. From this figure it can be seen that the $3072^3$ particles simulations, which are the most well converged ones, agree with the slope of $k^{2.5}$, and we conclude that the $k^{2.5}$ slope for small $k$-values is correct.

\begin{figure}
	\centering
	\includegraphics[width=1.00\textwidth]{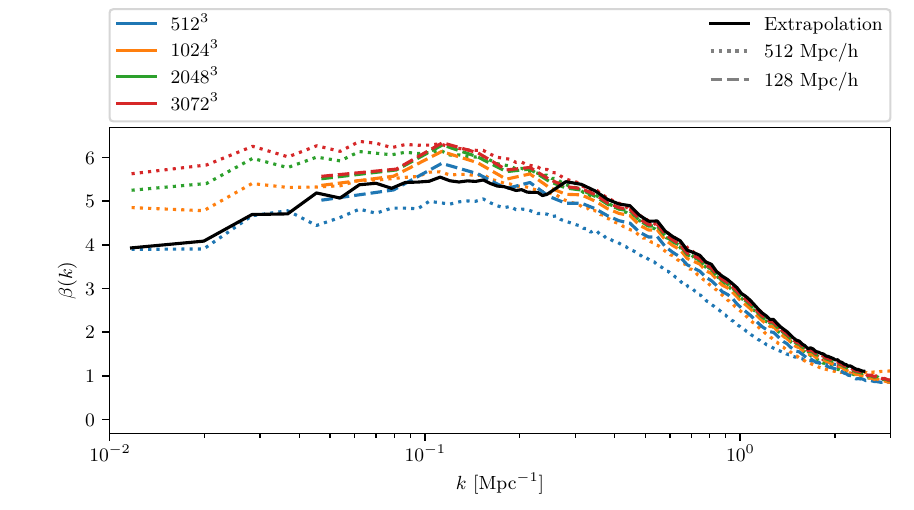}
	\caption{\label{fig:beta}$\beta(k)$ for the extrapolated and individual power spectra.}
\end{figure}

Using the extrapolated power spectra for $z=0$ and $z=1$, $\beta(k)$ is calculated using equation \eqref{eq:beta}, and the result is plotted in figure \ref{fig:beta} together with $\beta(k)$ calculated individually from the raw power spectra. $D_+(z=1)$ has been calculated theoretically with \class{} and has the value 0.61 for the cosmological model used here.

From figure \ref{fig:beta} we conclude that $\beta(k)$ asymptotes to a value of approximately 6 over a fairly wide range of $k$-values in the limit of very high resolution. Furthermore, we see that the value of $\beta(k)$ calculated from the extrapolated spectra is not stable for small $k$ because of the convergence issue at $z=1$ discussed above. Ref.~\cite{Pueblas:2008uv} found an estimated $\beta \simeq 7 \pm 0.3$, which is somewhat higher than our estimate. However, we note that their work is based on simulations at significantly lower resolution using a different mass assignment scheme, but also that they use results up to $z=3$ to fit their estimated value of $\beta$. These issues make it difficult to directly compare beyond noting that our results are roughly consistent with Ref.~\cite{Pueblas:2008uv}.

\subsection{Vorticity to divergence ratio and its convergence properties}

Finally, another interesting quantity to study is the ratio of power in the divergence and vorticity spectra, i.e.\ the splitting between infall and rotational motion in the velocity field.
Based on very simple considerations we expect that $P_\omega/P_\theta \to 0$ for $k \to 0$ because vorticity is only sourced by non-linear evolution. Conversely, for a fully isotropised velocity distribution one expects $P_\omega/P_\theta = 2$ based on the number of degrees of freedom in the two fields.

\begin{figure}
	\centering
	\includegraphics[width = 0.49\textwidth]{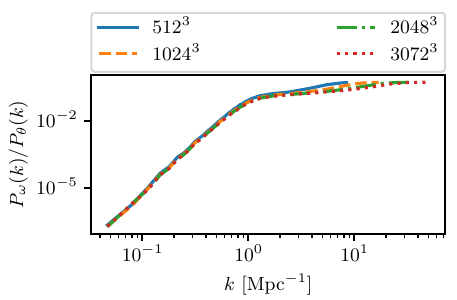}
	\includegraphics[width = 0.49\textwidth]{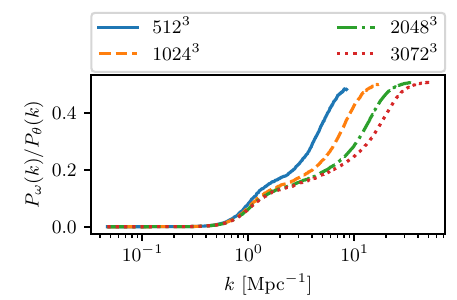}
	\caption{\label{fig:ratio}The ratio of power in the vorticity and divergence power spectra for different number of particles in the 128 Mpc/h box at $z = 0$. \textit{Left:} Log scale. \textit{Right:} Linear scale.}
\end{figure}

In figure \ref{fig:ratio} we show the ratio $P_\omega/P_\theta$ as a function of $k$ for simulations with different $X$, all in a 128 Mpc/h box and at $z=0$. As expected, the ratio remains extremely low until non-linearity is reached close to $k \sim 1/$Mpc. At this point there is a fairly sharp rise followed by a much more gradual rise in the ratio.
At higher $k$, an additional sharp rise is seen in the ratio. However, it is clear that this feature is a numerical artefact given that it  depends strongly on the resolution of the simulation. The most likely cause of this spurious rise in $P_\omega/P_\theta$ is the same interpolation error described in the previous subsection, and indeed it also becomes very prominent around $k \gtrsim k_\text{Nyquist}/6$.

Given our simulation constraints we can only claim good numerical convergence up to around $k \sim 5/$Mpc even in the $3072^3$ simulation. Beyond this limit we conjecture a continued slow rise in the ratio from around 0.2 at $k \sim 5$/Mpc, consistent with particle orbits being significantly more radial than tangential, except in the innermost parts of halos.
 
A number of previous studies have also calculated the ratio $P_\omega/P_\theta$ although typically at significantly lower resolution and often using different interpolation schemes which makes a direct comparison difficult. \cite{Pueblas:2008uv, Hahn:2014lca, Brandbyge:2017wyw} all find $P_\omega/P_\theta \sim 0.2-0.7$ in the non-linear regime and growing with $k$, roughly consistent with the present result, but somewhat higher (this is consistent with simulations in this work being run at significantly higher resolution).

\section{Discussion and conclusions}\label{sec:conclusion}

Based on the work presented in the above sections we can now draw a number of conclusions about the properties of the cosmic vorticity power spectrum.
First, we showed in section \ref{sec:pmvsp3m} that, because the cosmic velocity field is substantially more linear than the density field, it suffices to run PM simulations rather than P$^3$M simulations, even when precision requirements are at the 5\% level. This is true for both velocity divergence and vorticity even when the PM simulations have errors in the density power spectrum of several tens of percent on small scales.

Armed with this knowledge we proceeded to perform a number of high-resolution PM simulations to investigate convergence properties of the cosmic vorticity power spectrum. In order to avoid discretisation effects related to the possible presence of empty cells when using standard CIC interpolation, the spectra were calculated using the Delauney tesselation code DTFE, appropriately modified to handle very large simulations.
We discovered that while density and velocity divergence spectra have benevolent convergence properties, this is not the case for vorticity. The reason is that vorticity is not sourced at lowest order in perturbation theory and is therefore present only as a result of non-linear structure formation. This results in vorticity on large and intermediate scales being sourced by the emergence of non-linearities in density on smaller scales. This mode coupling makes it exceedingly hard to numerically converge the vorticity power spectrum in $N$-body simulations.

We found that numerical convergence within a {\it given} simulation box size is approximately controlled by the parameter $X\equiv N^{\frac{1}{3}} \frac{\text{Mpc}/h}{L_\text{box}}$. At $X = 4$ the power spectrum had approximately a 20\% error while $X = 24$ gave an approximate error of 5\%.
However, we also discovered that the dependence on the initial random seed used in the simulation can be much stronger for vorticity than for density or velocity divergence, especially in small simulation volumes. This effect most likely arises because vorticity in a small simulation box is primarily sourced by only a few very large structures going non-linear.
For the purpose of calculating the expected vorticity power spectrum we found that the strong dependence on initial random amplitudes and phases can be remedied by running simulations with amplitudes fixed to their expectation value (we found an additional minor improvement when running two simulations with opposite phases).
Finally, we also verified that results were robust against variations in the starting redshift of the simulations for $z_i  \geq 99$. 
Based on these findings we are confident that the cosmic vorticity power spectrum is calculable to approximately 5\% precision using the methods presented here. Given that the ``true'' vorticity power spectrum for a given cosmological model can be found using a small and well defined set of simulations combined with a simple extrapolation procedure it should be possible to investigate how vorticity depends on cosmological parameters without encountering prohibitive computational costs.

We also looked at the time dependence of the vorticity power spectrum. This was done by assuming that $P_\omega \propto (D_+(z))^{\beta(k)}$ and extracting the effective power-law index $\beta(k)$ from spectra at $z=1$ and $z=0$. Our finding is that $\beta(k)$ asymptotes to a value of approximately 6 in the limit of small $k$ and high numerical resolution. An interesting point for future work might be to run simulations for different models and over a wider range of redshifts and scales to establish the behaviour of $\beta(k)$.

Furthermore, we looked at the vorticity to divergence ratio and found that $P_\omega/P_\theta \sim 0.2$ at $k \sim 5$/Mpc, which is the highest $k$-value in our simulations with good numerical convergence of this quantity. This value is somewhat lower than what has been found in previous studies, but is consistent with our simulations being run at significantly higher resolution.

We finally note that the extremely large variance in vorticity seen in small box simulations also indicate that it will be very challenging to extract cosmological parameters from measurements of vorticity in small or medium sized volumes of the universe. \\

\noindent {\bf Reproducibility.}
We have used the publicly available code \concept{} available at \url{https://github.com/jmd-dk/concept/} to run the $N$-body simulations. To calculate the power spectra, we have used a modified version of the publicly available code \dtfe{} available at \url{https://github.com/AarhusCosmology/DTFE} together with the \textsc{Python} package \textsc{nbodykit} publicly available at \url{https://github.com/bccp/nbodykit}.

\section*{Acknowledgements}
We thank Román Scoccimarro, Mathias Garny, Francesca Lepori, and Ruth Durrer for valuable comments on the manuscript. We acknowledge the use of computing resources from the Centre for Scientific Computing Aarhus (CSCAA). CS and SH were supported by grant ``FLOWS'' from the Danish Research Council (FNU).
\appendix

\section{DTFE for large simulations}\label{sec:dtfemod}

For transparency and reproducibility, we briefly document modifications made to the \dtfe{} code in our publicly available fork\footnote{Link to the repository: \url{https://github.com/AarhusCosmology/DTFE}}.

\subsection{Handling \dtfe{} dependencies}

We have added key external dependencies as Git submodules to ensure easier setup and consistent builds across environments. The following submodules have been included:

\begin{description}
  \item[MPFR] The GNU Multiple Precision Floating-Point Reliable~\cite{mpfr} library provides precise and correctly-rounded floating-point arithmetic with arbitrary precision. It is used for high-precision numerical computations.

  \item[CGAL] The Computational Geometry Algorithms Library~\cite{cgal} offers reliable and efficient geometric algorithms. It is used by \dtfe{} for constructing the Delaunay tessellation.

  \item[GMP] The GNU Multiple Precision Arithmetic Library~\cite{gmp} is used for arbitrary-precision arithmetic on integers, rational numbers, and floating-point numbers. MPFR depends on GMP.

  \item[GSL] The GNU Scientific Library~\cite{gsl} provides a wide range of numerical routines, including integration, interpolation, and linear algebra. It supports scientific computing tasks within \dtfe{}.

  \item[Boost] A collection of portable C++ libraries~\cite{boost}. An attempt was initially made to replace Boost by STL equivalents, but \dtfe{} relies quite heavily on Boost so it was not easily done.

  \item[HDF5] The Hierarchical Data Format version 5~\cite{hdf5} is a data model and file format for storing large, complex datasets. It is the modern way of storing snapshots and is supported by \dtfe{}. However, due to various issues we ended up using the \textsc{gadget 2} binary format instead.
\end{description}

Including these libraries as submodules greatly simplifies the installation process. We made a few additional small changes as well, including fixing a 32-bit integer overflow, and an error that only occurred when reading a snapshot file with no particles.

\subsection{Using \dtfe{} for very large simulations}

\dtfe{} has a rather large memory footprint and is not MPI parallelised, which makes it tricky to use for very large simulations. For instance, computing the vorticity field in our $2048^3$ simulations was only barely possible on our 1.5TB memory node. There are a number of reasons for this. First off, \dtfe{} needs to hold the full snapshot in memory and also the full output data. For divergence and vorticity outputs, that is already 320GB. The Delaunay triangulation would take an additional $\sim 4$TB, although this part can be reduced by the partitioning functionality of \dtfe{}. We also found the memory requirements to scale with the number of Monte Carlo points used for the volume averaging and also to some degree the number of threads used for the computation.

Another problem is the uneven distribution of particles, which is a problem of small box-sizes in particular. In a $128\,\text{Mpc/h}$ box, a large fraction of the particles may become localised in a big cluster. Because \dtfe{} partitions particles in a regular grid, it cannot avoid having large variations of particle numbers across partitions. Thus, if we do a $(4, 4, 4)$-partitioning, \dtfe{} may work its way through the first 30 or so partitions before it hits the partition that triggers the out-of-memory error.

In order to use \dtfe{} for our large simulations we thus implemented the following two-step strategy:

\begin{enumerate}
\item Split the snapshots in $N_\text{split}^3$ fixed-size partitions with a fixed-width ghost zone of $5 L_\text{box}/N^{\frac{1}{3}}$. This operation is embarrassingly parallel across the $N_\text{sn}$ files in the multi-file snapshot, provided that we do not care about the total particle number from the header. (We could in principle update the headers very quickly after the splitting is complete, but \dtfe{} does not use that information anyway.) The result from this operation is $N_\text{sn} \times N_\text{split}^3$ files with a total size only slightly larger (due to ghost zones) than the original snapshot. We implemented this operation using a small \textsc{Python} script that was invoked $N_\text{sn}$ times for each redshift by a Slurm Array Job.
\item To analyse the output, we called \dtfe{} on each of the $N_\text{split}^3$ partitions using the \texttt{--region} flag of \dtfe{} and constructed a mosaic of outputs. This step was also implemented using the Slurm Array Job functionality.
\end{enumerate}

\begin{figure}[htbp]
    \centering
    \includegraphics[width=\textwidth]{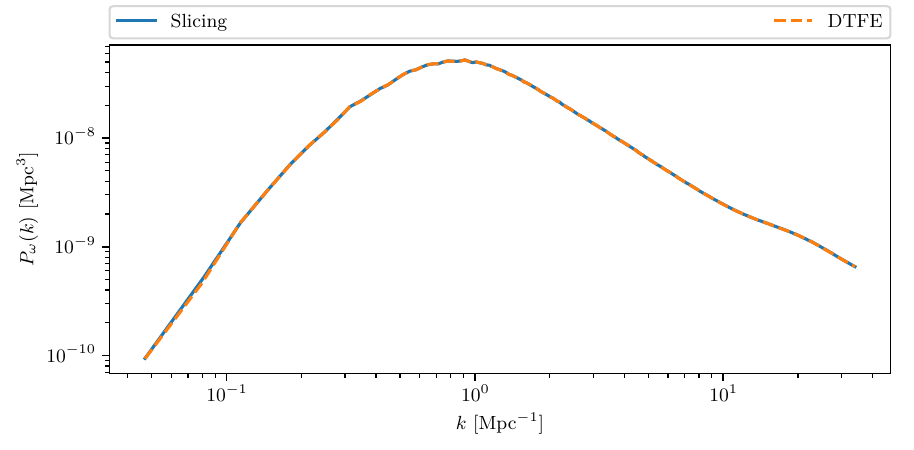}
    \caption{Spectra calculated for $N = 2048^3$ particles in a 128 Mpc/h box using standard \dtfe{} compared to the slicing method. From the figure it can clearly be seen that he slicing method does not leave any artefacts in the power spectrum.}
    \label{fig:slicing}
\end{figure}

We used $N_\text{split} = 4$ for $2048^3$ simulations and $N_\text{split} = 6$ for $3072^3$. Thus, we should expect $512^3$ particles on average in each partition. However, some partitions could have up to 4 times as many particles at low redshifts which would again trigger some out-of-memory errors. However, with our scheme it was straightforward to relaunch the failed runs with more memory. In figure~\ref{fig:slicing} we have verified that our slicing method does not generate artefacts in the spectra.

\subsection{Dependence on number of Monte Carlo samples}
\begin{figure}[htbp]
	\centering
	\includegraphics[width=\textwidth]{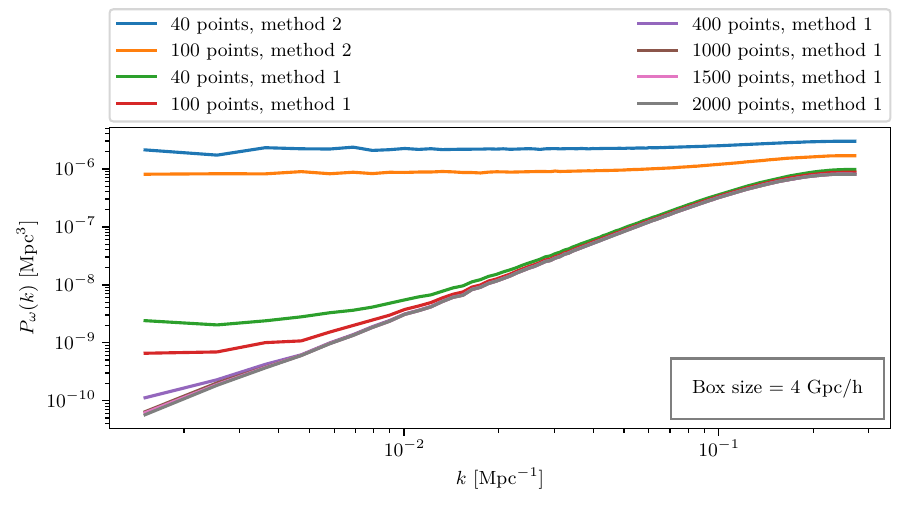}
	\includegraphics[width=\textwidth]{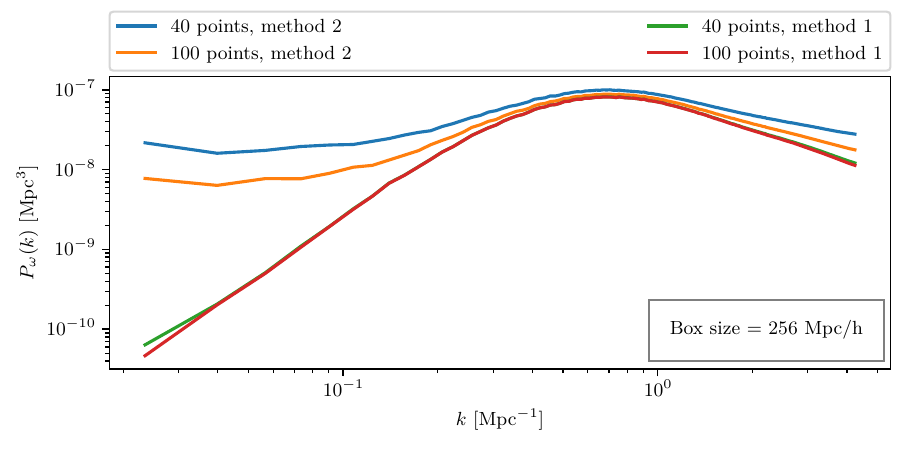}
	\caption{Vorticity power spectra for the same velocity field calculated using two different methods as well as different amounts of points used in the Monte Carlo sampling in a 4 Gpc/h box and a 256 Mpc/h box respectively. From this we decided to use method 1 with 1000 points for the Monte Carlo sampling.}
	\label{fig:samplesandmethods}
\end{figure}

To do the Delauney triangulation, \dtfe{} needs to sample points to use for the volume averaging, and it samples these using Monte Carlo sampling \cite{Cautun:2011gf}. To test how many points are needed, we ran \dtfe{} multiple times on the same snapshot using different numbers of points. The result can be seen in figure \ref{fig:samplesandmethods}. Furthermore, \dtfe{} also has two different methods to use when sampling, and figure \ref{fig:samplesandmethods} also shows the difference in running with each method respectively. From the figure we conclude that using method 1 together with 1000 points for the Monte Carlo sampling gives a good convergence even for the large 4 Gpc/h box when using \dtfe{}.

\bibliographystyle{utcaps}
\bibliography{vorticity2025}

\end{document}